\let\includefigures=\iftrue
\let\useblackboard=\iftrue
\newfam\black
\input harvmac

\noblackbox

\includefigures
\message{If you do not have epsf.tex (to include figures),}
\message{change the option at the top of the tex file.}
\input epsf
\def\figin{\epsfcheck\figin}\def\figins{\epsfcheck\figins}
\def\epsfcheck{\ifx\epsfbox\UnDeFiNeD
\message{(NO epsf.tex, FIGURES WILL BE IGNORED)}
\gdef\figin##1{\vskip2in}\gdef\figins##1{\hskip.5in}
\else\message{(FIGURES WILL BE INCLUDED)}%
\gdef\figin##1{##1}\gdef\figins##1{##1}\fi}
\def\DefWarn#1{}
\def\figinsert{\goodbreak\midinsert}
\def\ifig#1#2#3{\DefWarn#1\xdef#1{fig.~\the\figno}
\writedef{#1\leftbracket fig.\noexpand~\the\figno}%
\figinsert\figin{\centerline{#3}}\medskip\centerline{\vbox{
\baselineskip12pt\advance\hsize by -1truein
\noindent\footnotefont{\bf Fig.~\the\figno:} #2}}
\bigskip\endinsert\global\advance\figno by1}
\else
\def\ifig#1#2#3{\xdef#1{fig.~\the\figno}
\writedef{#1\leftbracket fig.\noexpand~\the\figno}%
\global\advance\figno by1}
\fi
%

\useblackboard
\message{If you do not have msbm (blackboard bold) fonts,}
\message{change the option at the top of the tex file.}
\font\blackboard=msbm10 scaled \magstep1
\font\blackboards=msbm7
\font\blackboardss=msbm5
\textfont\black=\blackboard
\scriptfont\black=\blackboards
\scriptscriptfont\black=\blackboardss

\else

\fi
%
\def\yboxit#1#2{\vbox{\hrule height #1 \hbox{\vrule width #1
\vbox{#2}\vrule width #1 }\hrule height #1 }}
\def\fillbox#1{\hbox to #1{\vbox to #1{\vfil}\hfil}}
\def\ybox{{\lower 1.3pt \yboxit{0.4pt}{\fillbox{8pt}}\hskip-0.2pt}}
%
%


\def\comments#1{}


\def\CL{{\cal L}}


\def\a{\alpha}

\def\ap{\alpha'}

\def\II{\relax{I\kern-.10em I}}

\def\IZ{\relax\ifmmode\mathchoice
{\hbox{\cmss Z\kern-.4em Z}}{\hbox{\cmss Z\kern-.4em Z}}
{\lower.9pt\hbox{\cmsss Z\kern-.4em Z}}
{\lower1.2pt\hbox{\cmsss Z\kern-.4em Z}}
\else{\cmss Z\kern-.4emZ}\fi}
\def\IB{\relax{\rm I\kern-.18em B}}
\def\IC{{\relax\hbox{$\inbar\kern-.3em{\rm C}$}}}
\def\ID{\relax{\rm I\kern-.18em D}}
\def\IE{\relax{\rm I\kern-.18em E}}
\def\IF{\relax{\rm I\kern-.18em F}}
\def\IG{\relax\hbox{$\inbar\kern-.3em{\rm G}$}}
\def\IGa{\relax\hbox{${\rm I}\kern-.18em\Gamma$}}
\def\IH{\relax{\rm I\kern-.18em H}}
\def\II{\relax{\rm I\kern-.18em I}}
\def\IK{\relax{\rm I\kern-.18em K}}
\def\IP{\relax{\rm I\kern-.18em P}}

%

\def\inbar{\,\vrule height1.5ex width.4pt depth0pt}

\font\cmss=cmss10 
\def\IR{\relax{\rm I\kern-.18em R}}

%


%

\def\lp10{\ell_p^{10}}
\def\lp11{\ell_p^{11}}
\def\R11{R_{11}}

\def\frac#1#2{{#1 \over #2}}

\def\imt{{\rm Im} \tau}


\def\uu{^}
\def\ll{_}
\def\a{\alpha}
\def\b{\beta}
\def\s{\sigma}

\def\m{\mu}
\def\n{\nu}

\def\sqd{^2}

\def\exp#1{{\rm exp}\{#1\}}

\def\t{\tau}


\def\1dag{^{1\dagger}}
\def\2dag{^{2\dagger}}

\def\R#1#2#3{{{R_{#1}}^{#2}}_{#3}}


\def\ttil{\tilde{\tau}}

\def\ie{{\it i.e.}}

\hyphenation{Di-men-sion-al}



\lref\bbst{A. Bernevig, J. Brodie, L. Susskind and N. Toumbas, ``How Bob Laughlin 
Tamed the Giant Graviton from Taub-NUT Space,'' hep-th/0010105.}

\lref\lenny{L. Susskind, ''The Quantum Hall Fluid and Non-Commutative Chern Simons Theory,''
hep-th/0101029.}

\lref\poly{A. Polychronakos, ``Quantum Hall states as matrix Chern-Simons theory,'' 
hep-th/0103013.}

\lref\hellerman{S. Hellerman and M. van Raamsdonk, ``Quantum Hall Physics = 
Noncommutative Field Theory,'' hep-th/0103179.}

\lref\hw{A. Hanany and E. Witten, ``Type IIB superstrings, BPS monopoles and three-dimensional 
gauge dynamics,''  hep-th/9611230.}

\lref\ggp{G. Gibbons, M. Green, M. Perry, ``Instantons 
and seven-branes in Type IIB superstring theory,'' hep-th/9511080.}

\lref\gz{M. Gaberdiel and B. Zwiebach, ``Exceptional groups from open strings,'' hep-th/9709013.}

\lref\gl{R. Gregory and R. Laflamme, ``Black strings and p-branes 
are unstable,'' hep-th/9301052.}

\lref\icantbelieveimdoingthis{C. Johnson, unpublished.}

\Title{\vbox{\baselineskip12pt\hbox{hep-th/0104010}
\hbox{SLAC-PUB-8808}\hbox{SU-ITP-01/15}}}
{\vbox{
\centerline{D(NA)-Branes}
}}
\bigskip
\bigskip
\centerline{Simeon Hellerman$^{1,2}$ and John
M${^{\underline{\rm c}}}$Greevy$^1$}
\bigskip
\centerline{$^{1}${\it Department of Physics, Stanford University,
Stanford, CA 94305}}
\smallskip
\centerline{$^{2}${\it SLAC Theory Group, MS 81, PO Box 4349,
Stanford, CA 94309}}
\bigskip
\bigskip
\noindent
We engineer a configuration of branes in type IIB string theory 
whose mechanical structure 
is that of a DNA molecule.
We obtain it by considering a T-dual description of the 
quantum Hall soliton.
Using a probe analysis, we investigate the
dynamics of the system and show that it is stable against 
radial perturbations.
We exercise a certain amount of restraint in discussing applications
to biophysics. 

\bigskip
\Date{April 1, 2001}

\newsec{Introduction}

String theory -- at present the only known consistent theory of quantum
gravity -- has undergone several major changes during its lifetime in terms of
the way in which physicists look at it,
and what they expect from it.  Originally it was
viewed as a possible theory of strong interactions, although the
ineluctable 
appearance of a massless spin-2 state in its spectrum dampened 
the enthusiasm for the idea.
When this phenomenon was later reinterpreted as a feature rather than a bug,
with the spin-2 state incarnate as a graviton rather than an anomalously
light hadron, the theory garnered a reputation as the leading
candidate for a theory incorporating the phenomenon of gravity 
into quantum mechanics.  The
discovery of supersymmetry, a cornucopia of compactifications to four
dimensions, and quasi-realistic gauge groups and matter content in the
string framework only served to solidify the hegemony of strings over the
field of theoretical high-energy physics.

Much of the recent excitement in string theory 
has come from an (often nonperturbative) understanding of 
many different backgrounds of the theory and 
the relations between them.
The protean character of quantum states under the duality group of
string theory is a source of much of the continuing fascination with the
theory, and of the general sense that it has many depths yet to be explored.
Indeed it seems that just about any quantum theory imaginable can be
obtained as a (low-energy or other) limit of the dynamics of the theory.
Its shape-shifting capacity appears unlimited.

Most recently, the theory has been investigated because of the fact that
one of its backgrounds seems to be able to reproduce the quantum Hall
effect at low energies.  In fact two logically separate dualities relating
the Hall system to other theories of interest have emerged recently.  The
first \refs{\bbst} is a direct embedding of the system in string theory.  The
second \refs{\lenny} is an equivalence to a noncommutative gauge theory
whose relation to string theory is less direct.  This equivalence was
recently sharpened in \refs{\poly, \hellerman}.

In this paper we describe a closely related system, obtained by considering
the T-dual of (a nonabelian version of) the system in \refs{\bbst}.
We show that the T-dual is a double helix, with fundamental
string rungs connecting the two helices.

The plan of the paper is as follows.  In \S2, we briefly review
the construction of the quantum Hall soliton, and motivate an
examination of the D(NA)-brane system by considering its T-dual.  In
\S3 we use the Dirac-Born-Infeld/Wess-Zumino action on the lighter branes to
find the equilibrium radius of the helix.  In \S4 we describe the
low-energy dynamics of the system. 
In \S5 we conclude by discussing the limitations of our calculation
and ways in which it might be improved.

\newsec{The quantum Hall soliton and its DNA dual}

We briefly review the construction {\refs{\bbst}} of a quantum Hall-like
system in type IIA string theory.

Begin with a set of $k$ sixbranes, spatially extended in the $x^4,\cdots,
x\uu 9$ directions, located at transverse position $x_1 = x_2 = x_3 = 0$.
Surrounding these sixbranes we place a set of $n$ coincident twobranes --
that is, twobranes whose location is defined by the sphere $x_1^2 +
x_2^2 + x_3^2 = R^2_0$, with $R_0$, the radius of the sphere,
determined by the dynamics and specified later.
When we say 'surrounding' we mean that the twobranes literally cannot
be moved off to infinity without intersecting the sixbranes.

It is known \refs{\hw} that for such nontrivially 'linked' configurations
of D-branes in type II string theories there is an effect that causes
fundamental strings to connect the two linked objects.  Referred to as
the 'Hanany-Witten' effect after its discoverers, this feature
of brane dynamics can be understood from
many different points of view, {\it c.f. e.g.} \gz. 

In order to stabilize the configuration one can dissolve $n$ zerobranes in
the set of twobranes.  There is a repulsive force between
zerobranes and sixbranes at long distances, and so the system seeks
its equilibrium radius, calculated in \refs{\bbst} to be 
$R_0 =(\pi k'/n)^{2/3} l_s/2 $.

The authors of \refs{\bbst} argued that the dynamics of this system were
closely related to those of the quantum Hall effect, in that the string
endpoints appeared in the gauge theory dynamics as charges in the fundamental
representation (in \refs{\bbst} only a single twobrane was considered, and
so the string endpoints were simply electric charges) and the dissolved
zerobranes were units of magnetic flux.  For a single twobrane the 'filling
fraction' of the system was simply $k / k^\prime $, the ratio of the
density of electrons to the density of flux.

\ifig\tdualA{The periodic array of quantum Hall solitons before 
the Gregory-Laflamme transition.}
{\epsfxsize3.5in\epsfbox{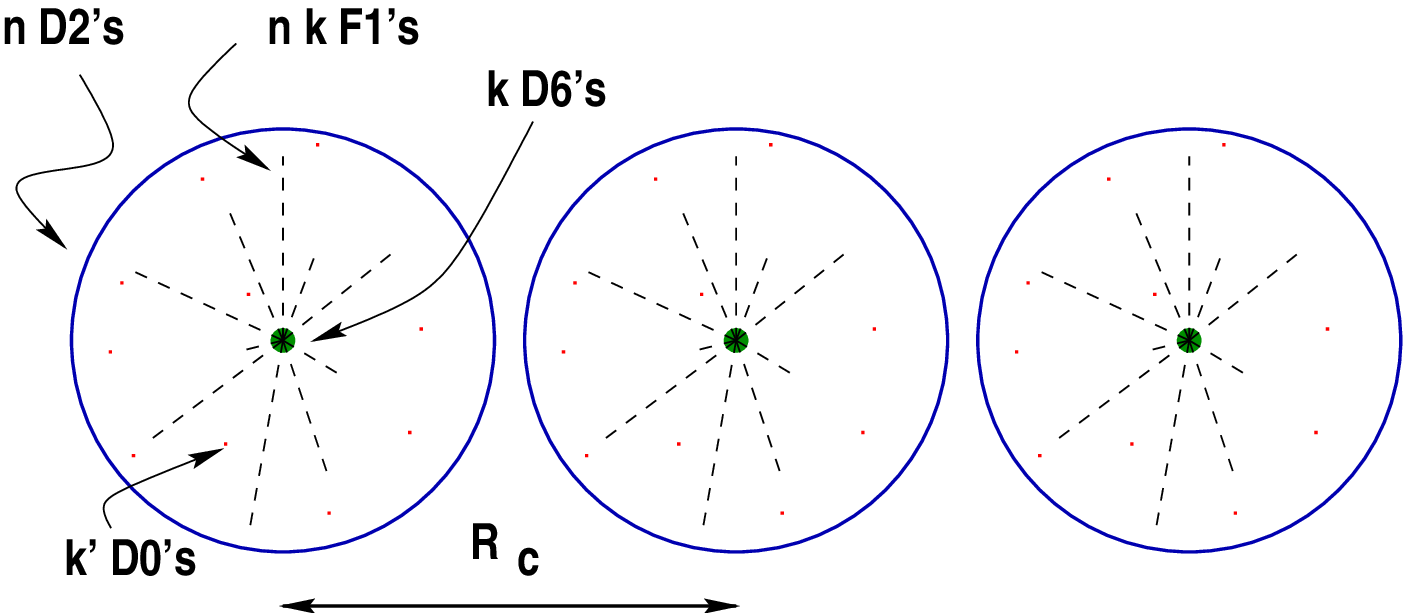}}
Now suppose one compactifies the system along the $x_3$ direction with
radius $R_c$.
For $R_c >> R_0$ we
can think of the periodically identified configuration as an infinite
array of twobrane spheres surrounding an array of sixbranes as in 
\tdualA\foot{Strictly
speaking as soon as one compactifies transverse to the sixbranes
one runs into trouble because the fields generated by the branes
grow logarithmically in the remaining transverse directions
and there is a conical deficit at infinity.
For now we assume we can regulate this problem, but we will discuss the point
in more detail further on.}.  
We expect, however, that as we reduce the
compactification radius until $R_c \sim R_0$, the system will develop an
instability (closely related to the Gregory-Laflamme instability \gl) towards
a merger of the twobranes into a cylinder, periodically identified
in the compact direction.  
\ifig\tdualB{After the Gregory-Laflamme transition.}
{\epsfxsize3.1in\epsfbox{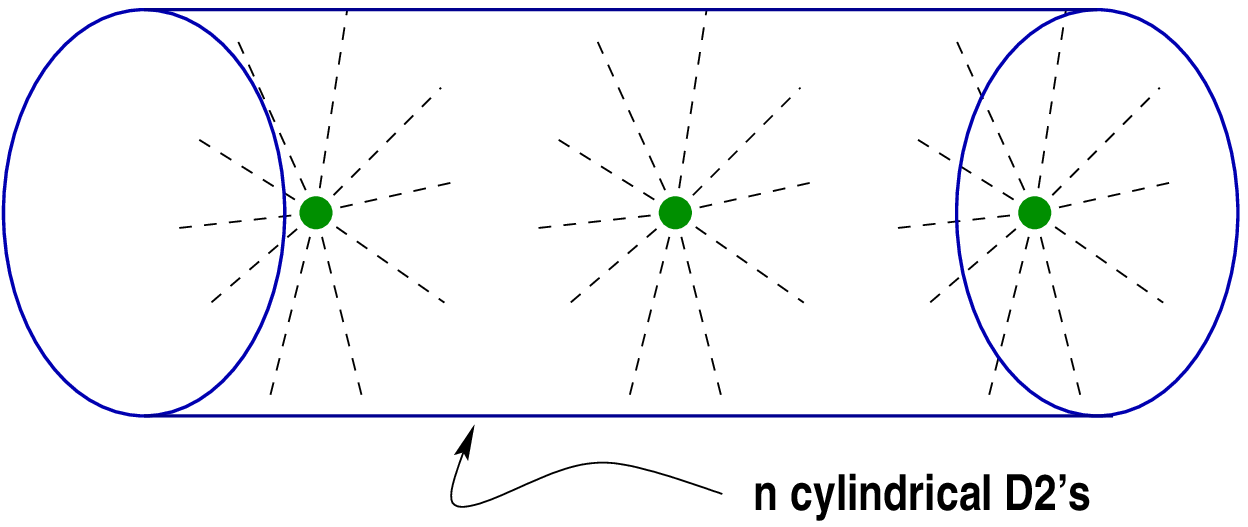}}

Now we perform $T$-duality along the $x_3$ direction.  Under this duality:

$\bullet$ The $k^\prime$ zerobranes become $D1$-branes, extended in the $x^3$
direction.

$\bullet$ The $k$
sixbranes become $D7$-branes, extended in the $x_3, x_4, \cdots,
x_9$ directions.

$\bullet$ The $n$ cylindrical twobranes become circular onebranes in the
$x_1, x_2$ plane.

$\bullet$ The $n k $ fundamental strings stay as they are; now they are
stretched between the sevenbranes and the circular D-strings surrounding
them.

\ifig\tdualC{The unstable D-string configuration.}
{\epsfxsize3.1in\epsfbox{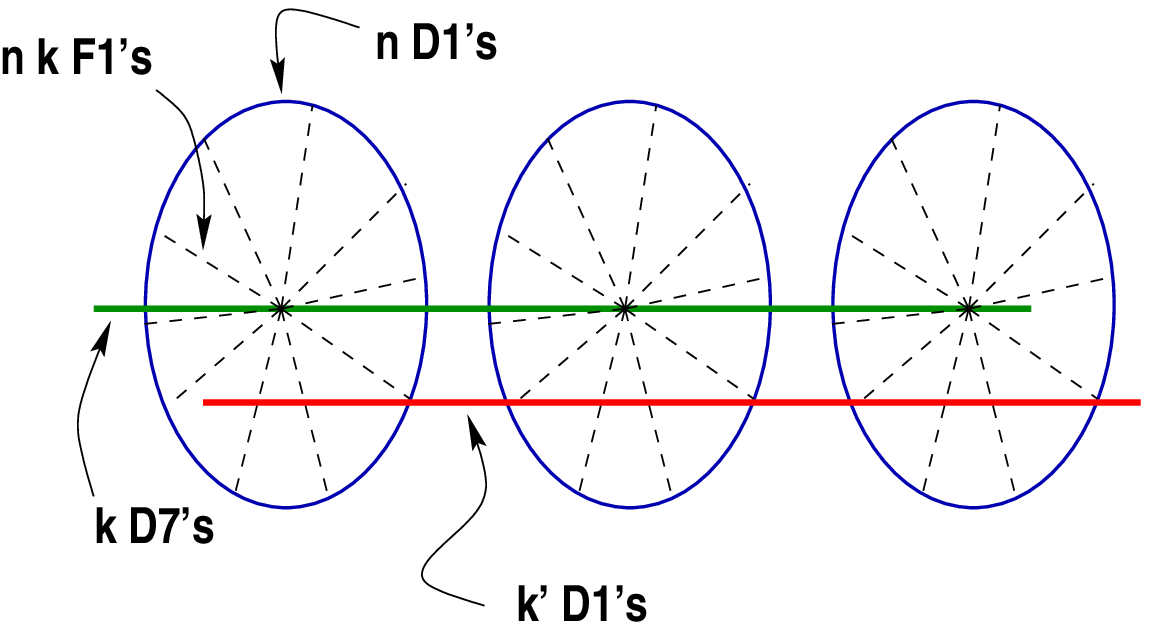}}
In fact this description 
is not very accurate; it is merely a cartoon which illustrates the way in
which the charges transform under $T$-duality.  The configuration shown
in \tdualC\ would more accurately describe 
the $T$-dual of
the unstable system in which the zerobranes are present but not dissolved in
the twobranes.  The correct, (meta)stable configuration in the original
picture is the one in which the zerobranes and twobranes form a bound state
in which the zerobranes give up almost all their rest energy. 
The correct $T$-dual picture is one in which the circular onebranes
'bind' to the straight onebranes by forming a coil -- the preference of the
system for the bound state is a consequence of the Pythagorean theorem.
\ifig\tdualD{How Watson and Crick tamed the 
D1-D7 system.}
{\epsfxsize4.0in\epsfbox{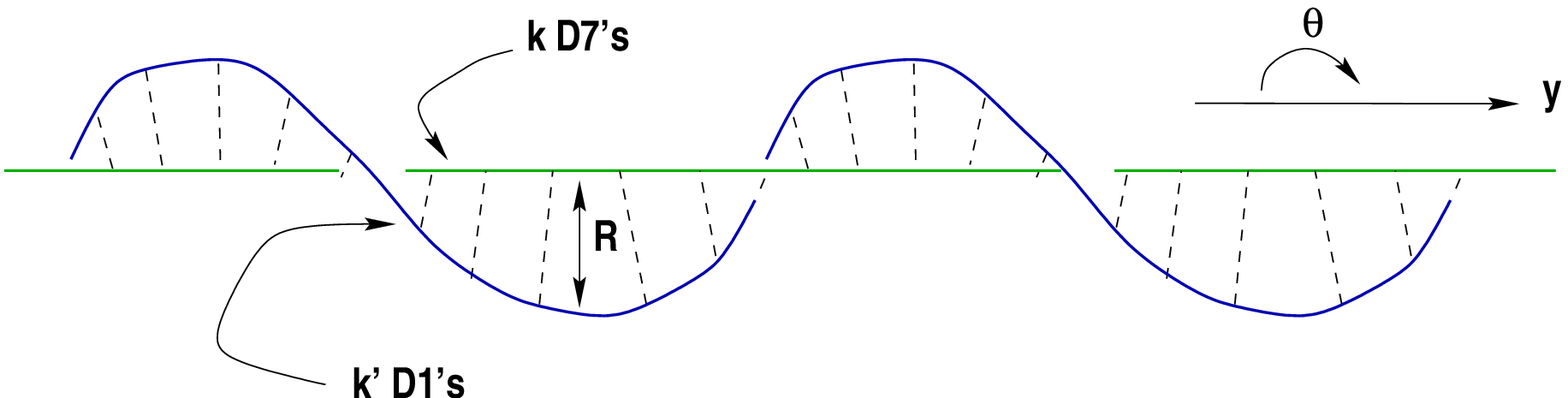}}

Note that the picture above is most accurate if the ratio $n/k^\prime$ is
large,
a limit opposite that of $\refs{\bbst}$, in which $n=1$.
For $n/k^\prime >>1$ (so that the number of coils is large) and $R_c / \alpha
^\prime <<1$
(so that the T-dual radius is large) we really can consider an infinitely
extended D(NA)-brane.  Specifically, in order to keep fixed the number
of coils per unit length of a bundle of $k^\prime$ $D1$-branes, we need to
take a large-$n$ limit of the nonabelian quantum Hall soliton:
$$
\eqalign{
n &\to \infty
\cr
R_c &\to 0 }
$$
holding fixed the natural quantities in the $T-$dual picture:
\eqn\scaling{
\eqalign{
k^\prime = \hbox{\rm number of onebranes in a bunch} \hskip 1 in &{\rm (fixed)}
\cr
{{n R_c}\over {k^\prime \alpha^\prime}} = \hbox{\rm number of
coils per unit length}
\hskip 1 in &{\rm (fixed)}
\cr
k = \hbox{\rm number of sevenbranes in a bunch} \hskip 1 in &{\rm (fixed)}
}
}
We also keep fixed the coupling in the resulting T-dual 
theory (this means we have to scale the original coupling).
Note that the number of fundamental strings per coil is 
$${ \hbox{\rm number of F-strings per $R_c$} 
\over \hbox{\rm number of coils per $R_c$}} = { nk \over n/k'} = k k'.$$

Consider for a moment taking $k'=0$. 
In that case, the D-strings 
would not coil in the $y$-direction at all, and would close onto 
themselves.  Recall that when a fundamental string ends 
on a D-string, the D-string carries away the F-string charge 
in one direction (as worldline flux).  
\ifig\monodrome{
Checking consistency 
between sevenbrane monodromy and charge conservation.}
{\epsfxsize2.5in\epsfbox{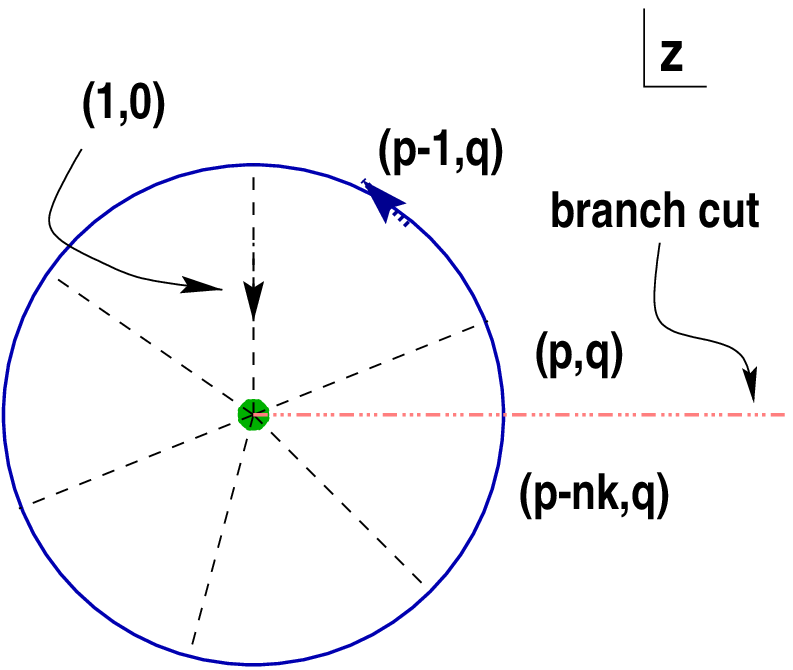}}
In order to have charge 
conservation when the D-string closes, the monodromy action 
of the D7-branes on the $(p,q)$-string charges 
must cancel off this accumulated F-string charge.
If we start off with a $(p,q)$ string, and go 
around the k D7 branes which emit $nk$ F-string 
spokes, we must have (in the notation of \gz)
$$
\left ( \matrix{ p \cr q } \right) = 
M_{[k, 0]} \left ( \matrix{ p - n k\cr q } \right) = 
\left (\matrix{ 1 & k \cr 0 & 1 } \right) 
\left (\matrix{ p - n k \cr q } \right) = 
\left (\matrix{ p - n k + k q \cr  q }\right),
$$
where 
$M_{[p',q']} = 
\left (\matrix{ 1 - p'q' & p'^2 \cr -q'^2 & 1 + p' q' } \right ) $ 
is the monodromy 
experienced by the 
charge lattice of $(p,q)$ strings 
in traversing the branch cut 
of a $(p', q')$ 7-brane \gz.
Therefore we see that we must have $q = n$ units 
of D-string charge, and that the number $p$ of 
units of F-string charge is arbitrary.
Once we take $k'$ nonzero, the string moves 
in the $y$ direction as it coils and 
this condition 
illuminates 
the Hanany-Witten effect from the 
F-theory point of view.

\bigskip
\noindent
{\it The Double Helix}

Finally, 
consider what will happen if we 
compactify the six dimensions $x_4,\cdots x_9$ on a $T_6$ of total
volume $V_6$ (whose individual dimensions will not figure in the discussion), 
giving the wrapped sevenbranes a finite tension.  This will 
cause them to coil up in response to the pull of the 
stretched fundamental strings, and 
the resulting object will be a double helix.
The tension of the effective one-dimensional object in $3+1$ dimensions
will be $V_6 / (\alpha^{\prime 3})$ times that of a onebrane; we will let
$V_6$ be large enough that we can treat the bundle of sevenbranes as
a heavy, fixed background in which the onebranes move as probes.

Before analyzing the stability of the system, there is one final 
subtlety to be dodged.  When one considers an infinitely extended
one-dimensional object in three spatial dimensions, one must deal somehow
with the fact that objects of spatial codimension two have rather dramatic
behavior when they couple to massless fields, particularly to the metric.
The sevenbranes and onebranes are sources for the $RR$ zero-form and
two-form potentials, respectively, which grow logarithmically rather
than falling off at large distance from the source.
Also the coupling to the metric creates a conical deficit at
infinity.  Furthermore if one tries to add too many sevenbranes and onebranes,
one will actually drive the deficit beyond $360^{\rm o}$.
Finally both the onebranes and sevenbranes source the dilaton and other moduli,
which grow
logarithmically away from the branes, changing the way in which one defines
the adjustable coupling for this background.

While there may be ways to regulate the long-distance fields of an infinite
string consistently, we take a strictly pragmatic approach in this paper
and instead of an infinite D(NA)-string, we consider an object 
with a different ``tertiary structure''.

Wind the helix around
in a large loop with
a radius far larger
than the string scale, the radius of the helix, or the scale on which the 
helix is coiled.  
\ifig\loopy{
Our regulator.}
{\epsfxsize2.5in\epsfbox{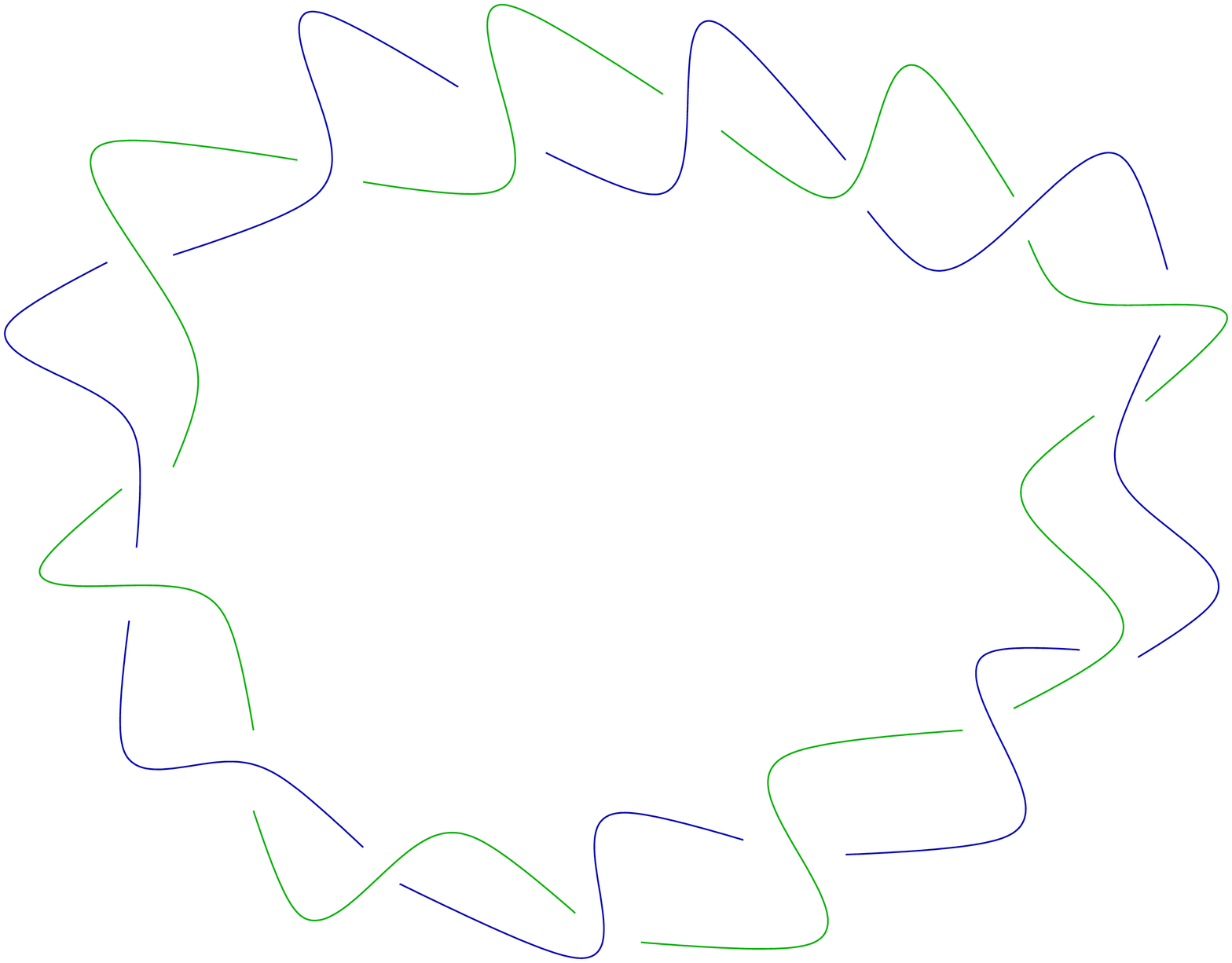}}
Obviously, if we make it big enough, the timescale for 
its collapse will be much longer than those relevant to the 
``secondary structure''.
Among other simplifying features, this regulated
configuration has unambiguous
expectation values
of the dilaton and other moduli at infinity, although 
the dynamics on the onebranes will decouple from the asymptotic dilaton.  

\newsec{Probe analysis and low-energy dynamics}

The complexified dilaton is 
$$ \tau(z) = \chi + i e^{-\Phi} .$$
The Einstein-frame metric for sevenbranes with worldvolume along 
$t, y, x_4, \dots x_9$ is \ggp\
$$ ds_{{\rm ein}}^2 
= - dt^2 + dy^2 + \sum_{i=4}^9 dx_i^2 + \Omega^2 dz d\bar z ,$$
where $\Omega$ is a function of $z$ defined below.
The string-frame metric is 
$ ds^2 = {1 \over \sqrt{ {\rm Im }\tau}} ds_{{\rm ein}}^2$.
We work in units where $\ap = 1$.  Let $z = r e^{i \theta}$.  

We take the near-brane solution for a stack 
of k D7-branes, for which the complexified 
dilaton takes the form
$$\tau(z) 
=j^{-1}( -bz^k) 
\simeq {k \over 2 \pi i} \ln b z.$$
This approximation is valid at weak coupling, 
\ie\ when $\tau_2 \equiv \imt$ is big.
$b$ is a parameter of the solution 
which plays the role of a dilaton modulus.
The metric function $\Omega$ takes the form
$$ \Omega^2 = \tau_2
| \eta(\tau)^2 \prod_i (z - z_i)^{-{1 \over 12}} |^2 
\simeq \tau_2 | (bz)^{2 k\over 24} z^{-{k \over 12}} |^2 
\simeq -{k \over 2\pi} b^{k \over 6} \ln ( b R) 
\equiv -{k \over 2 \pi \gamma^2} \ln (b R) , 
$$
where $\eta$ is the Dedekind eta function.

Parametrize the worldvolume of the helix by a spatial coordinate 
$\sigma$ and choose the embedding 
$$ y = \alpha \sigma, ~~ z = R e^{i\sigma} ,$$
so that $\theta = \sigma ~ {\rm mod} ~ 2 \pi$.
$\alpha$ is the inverse number of coils per unit length as defined by 
the scaling limit \scaling.

Then, ignoring kinetic terms because 
at the moment we are just interested in finding the 
potential for $R$, 
the induced $E = G + F$ (using the string-frame metric) is 
$$ 
\eqalign{E = (G+F)_{\a\b} 
d\sigma_\alpha d\sigma_\beta &= 
(G\uu{\m\n} \del_\alpha X_\mu \del_\beta X^\nu + F_{\a\b})d\s\uu\a d\s\uu\b \cr
&= 
\left ( \matrix{ dt, d\sigma } \right) 
\left (\matrix{-{1 \over \sqrt{\tau_2}} & \xi \cr -\xi & 
{1 \over \sqrt{\tau_2}}(R^2 \Omega^2 + \alpha^2)} \right ) 
	\left (\matrix{dt \cr
		       d\sigma} \right ),}
$$
where $\xi = F_{0 \sigma}$ is the electric field 
along the helix.  
So we have 
$$-{\rm det} E = e^\Phi (\Omega^2 R^2 + \alpha^2) - \xi^2 .$$

Let us discuss the form of $\xi$.  The sources for the 
field strength $\xi$ are the endpoints of the fundamental 
strings on the D-string, and the background axion gradient:

$$ \CL_{\xi} \sim 
{1 \over g_{YM}^2} \xi^2 - 
A_0 \left ( {1 \over k'} \sum_i \delta({\rm string~end}_i) - 
			       \del_1 \chi \right),
$$
where $g_{YM}^2 = e^{\Phi}$ and $\xi d\sigma dt  = dA$.  The factor of ${1 \over k'}$ 
multiplying the string source arises because a single string-end ending 
on a clump of D-strings sources the 
trace of $\xi$ with unit strength.
We find that the solution should be

$$ 
\eqalign{ \xi &\sim e^{\Phi} \left( {k \sigma \over 2 \pi} - 
{2\pi \over k'} \sum_i \Theta({\rm string~end}_i) \right) - \xi_0 \cr
&= {-2 \pi \over  \ln bR } 
\left( \sigma - {2 \pi \over k k'} \sum_i \Theta({\rm string~end}_i) \right) - \xi_0.
}
$$  
$\xi_0$ is a background electric 
field determined by the worldline theta-angle, 
which is in turn related to the bulk RR axion. 
As we will explain below, the worldline theta-angle becomes 
dynamical and chooses $\xi_0$ to make the average electric field vanish.

The electric field on the D-strings has the form of a sawtooth which reaches 
its maximum when it reaches a string end which then discharges it.
However, in the following, we take a average field approximation, where 
we replace $\xi$ by its spatial average value.  This approximation 
is justified by the fact that the deviation of $\xi$ from its average 
value is of order ${ 1 \over k k'}$ which we take to be small.  

Plugging into the Dirac-Born-Infeld plus Wess-Zumino 
probe action, 
$$ S = \int dtd\sigma \CL = k' \int dt d\sigma \left( e^{-\Phi} 
\sqrt{ - {\rm det} E } 
+ e^{-(F + B)} \sum_p C^{(RR)} \right), 
$$
(the $k'$ is out front because there are $k'$ D-strings)
we find
\eqn\potDstr{ \CL =  
k' \sqrt{ 
\tau_2
\left( 
({ R^2 \tau_2 \over \gamma^2} ) + \alpha^2 - \tau_2 \xi^2 \right)}
+ {k k' \sigma \over 2 \pi} \xi .}

We must also include the force on the D-string from the 
tension of the attached fundamental strings.  
These contribute an energy proportional to 
their length in the string frame metric, which is 
in turn equal to their coordinate length, over $\gamma$.  Averaging 
over $\sigma$, we find that they contribute 
a linear potential, 
$$V_{{\rm strings }} = {k' k \over 2 \pi \alpha' \gamma} R.$$

In the average field approximation, 
a convenient way to write the potential for $x = bR$ is 
\eqn\potent{
V = {k k'\over 2 \pi \gamma b} \left( x + \sqrt{ x^2 \ln^2 x - c \ln x } \right) 
}
where $c \equiv 2\pi \alpha^2 b^2 \gamma^2 /k$, a dimensionless parameter.
For small but nonzero $c$, this function has a minimum 
at $x_E$ defined 
by the transcendental equation
$$ c = 2 x_E^2 \left( \ln^2  x_E + \sqrt{2} |\ln x_E|^{3/2} \right) .$$
There is a critical value of $c \simeq 0.6689$ 
above which there is no minimum, and the potential just slopes 
off toward infinity.



\bigskip
\noindent
{\it Issues raised by this calculation}

\item{1.} Obviously we have only considered the dynamics of 
a single mode of the helix, albeit the 
most obvious candidate for an instability.  
In the next section, we consider some other modes.

\item{2.} The sevenbrane geometry which we are probing 
is singular if $k \neq 24$.  Further, there is a range of $r$'s 
($ r > 1/b$) for which 
our near-brane approximation breaks down and 
the dilaton, $\Phi$, is apparently imaginary.  
However, 
the minimum we found above 
lies within the region where 
our approximation makes sense.  
\ifig\min{The potential as a function of $x$ in units of 
${k k' \over 2 \pi \gamma b}$ for $c = 0.5$.
The gravity 
solution goes stupid around $x \sim 1$.}
{\epsfxsize2.5in\epsfbox{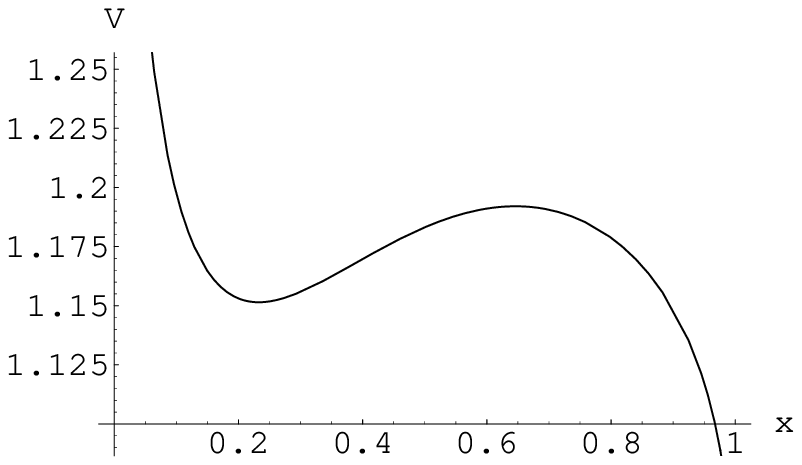}}

\item{3.} The DBI analysis is only valid if the brane 
worldvolume is weakly curved, and its field strengths are 
slowly varying; this is the case for our probe, except for the step-function
discontinuities due to fundamental string sources.

\item{4.} The value of the dilaton 
at the D-strings,
$$ e^{\Phi(R_E)} = { 2 \pi \over  k |\ln x_E |} $$
can be made parametrically small 
by taking $k$ large and $\alpha^2$ large fixing $R_E, c, b$.

\item{5.} If we take $c$ smoothly to zero, the minimum we found 
above moves closer to the sevenbranes.  
In this limit, the 1-7 strings become light, but 
the coupling to the 7-7 strings can be 
kept small by making $V_6$ large.
Since $c \propto \alpha^2$, 
this is 
consistent with the T-dual fact that in the absence 
of zerobranes, the quantum Hall soliton collapses.

\item{6.} Observe in \min\ that our potential 
has a maximum, and slopes downward far from the sevenbranes.  
When $c$ is small, this maximum is inside the 
region of validity of our approximation.  This 
signals that our equilibrium is perhaps only 
metastable, as is the quantum Hall soliton.

\newsec{Low-energy theory on the strings}

What can we say about the effective dynamics on the worldline of the helix?
The low-energy effective excitations are as follows:

$\bullet$
There are six compact scalars from the transverse $T^6$.  
The $T^6$ is large and the sevenbranes are distant, 
and we assume in what follows that these modes decouple.
There is also a seventh Goldstone mode, corresponding to translation
in the axial direction, which is likely to decouple as well.

$\bullet$ We found that there is a (stable) equilibrium value for the radion
field, $R$, that measures the coordinate distance between the onebranes and
sevenbranes; the radion is massive.

$\bullet$ There is a charge displacement wave mode, $D$, which 
is  massive.  
In the presence of a neutralizing background charge (provided by the axion
gradient), the charged string endpoints are bound to their equilibrium
positions with a linear restoring force.

$\bullet$ There is a
mode that rotates the entire onebrane around the sevenbranes, leaving the
$F$-strings at fixed axial position.  This 'turnon'\foot{Since the word
'roton' already has a standard usage in
condensed matter physics, we were left with little choice in the matter.} field
$T$ is
not an independent mode; it can be compensated by a combined charge
displacement and axial shift.  This corresponds with the fact that
in the presence of a background axion that winds $k$ times around the
sevenbranes, the turnon is in effect a dynamical theta angle
in $1+1$ dimensions;
giving $T$ a vev produces a vacuum energy proportional to $T\sqd$ which matches
the energy from the corresponding charge displacement.

$\bullet$ There are modes corresponding to fragmentation of the onebrane into
constituent onebranes.
There are also fragmentation modes along the $T^6$ directions.
Understanding the fragmentation modes would require a more careful
treatment of the nonabelian dynamics than we attempt here.  In the probe
analysis, then, we will set $k^\prime \equiv 1$.  However abelian and
nonabelian coulomb forces are quite similar in $1+1$ dimensions, and so
we believe the qualitative picture may be similar when $k^\prime
\neq 1$.

$\bullet$
There is no independent 'unwindon' -- that is, the
mode which uncoils the D-strings is essentially a linearly rising mode
of the turnon field.

\subsec{The full potential}

In this subsection, we perform a refinement
of the calculation of \S3, for the case $k^\prime = 1$.

Let us fix $y\equiv \a\s, X\uu 0\equiv \t$
identically as a helical analog of static gauge
for the reparametrization invariance of the DBI action.  In order to find
the roton and turnon potential, we write
\eqn\dbi{
z \equiv R(\s) \exp{i (\s + T(\s)) } 
}

First we set the charge density to its equilibrium value, and find a potential
for $R$ and $T$ alone.  We apply the 
approximation in which the fundamental string endpoints are continuously
and uniformly distributed so as to cancel the background charge density coming
from the axion gradient.  The electric field on the onebrane worldvolume is
constant, and equal to $\xi$.  The energy is at a minimum when $\xi$
vanishes.  In this system the $\theta$ angle is dynamical, so the electric
field can relax to zero by changing the value of the turnon.

Substituting this parametrization into the probe action we expand to zeroth
order in derivatives, for the full Lagrangian:
\eqn\dbipotone{
\CL (R, T)  =  e^{- \Phi(R) }\sqrt{\alpha\sqd + R\sqd \Omega\sqd (R)-
\xi\sqd} + T \xi - {k \over 2 \pi \gamma } R.
}
Notice that the turnon field only appears in the WZ term, 
coupling as a Peccei-Quinn axion.
The
background electric field is determined by the 
effective theta angle:
$$
\xi = {{\ttil\ll 1}\over{\sqrt{\ttil_2}|\ttil(R, T)|}} \cdot \sqrt 
{\a\sqd + R\sqd \Omega \sqd (R) }
$$
where $\ttil \equiv T + i \cdot
e^{- \Phi(R)}$ is the effective gauge coupling.  
Note that $\xi$ has an interpretation as a density of dissolved 
Wick-rotated D$(-1)$-branes.
Integrating out $\xi$ in the manner above is analogous to the 
``dilute instanton gas'' approximation in four-dimensional 
axion physics.

Plugging this back into the Lagrangian we obtain
\eqn\fullpot{
V = - e^{\Phi(R) / 2}
\sqrt {\a\sqd + R\sqd \Omega\sqd(R)} \cdot |\ttil(R,T)| + {k \over 2 \pi \gamma} R.
}

Having done this calculation, it is easy to take account of the
motions of the charged string endpoints.  A charge distribution
couples to the gauge field as
\eqn\chargedistrib{
L \to L + \rho(\sigma) A_0 (\sigma).
}
However for small motions of the charges from their equilibrium positions,
the charge density is given by the {\it inhomogeneity} of the charge displacement
field $D(\sigma)$; that is $\rho (\sigma) = D^\prime(\sigma)$, so
integrating by parts we find the interaction term
\eqn\chargedistribtwo{
- D(\sigma) \xi.
}

So $T$ and $D$ enter the action only through their difference.  To find
the full potential for all the fields, including the charge displacement,
simply substitute $T-D$ for $T$ in the expression \fullpot.
This expresses the fact that any uniform displacement of the charges from their
equilibrium positions can be compensated by a rotation of the helix.

In the end we find a $(1+1)$-dimensional system of three 
coupled fields: the charge density of string-ends, the 
turnon field, and the radion.  In the absence of 
a charge-clumping instability, we expect that the radion field decouples.  
The resulting system seems to form a Wigner crystal.
It will be interesting 
to learn more about the D(NA)-brane system, 
particularly when $k^\prime \neq 1$.

\newsec{Conclusions}

We have shown that the quantum Hall soliton has a certain limit in which
it is naturally viewed via T-duality 
as a molecule of DNA.  Though there is still much we do not understand about
the D(NA)-brane system, the dynamics are those of point
charges in a neutralizing background.  In addition to the Goldstone modes,
the theory on the strand contains a
worldvolume axion, the turnon, and a charge displacement field, one
combination of which is massless.

Our computation is incomplete in the following ways:

$\bullet$ 
In our low-energy analysis, we have 
merely determined which modes have nonzero mass.
In order to compute physical masses of fields, one would 
need to compute their kinetic terms.  

$\bullet$ 
Where uncertain, we have given the benefit of the doubt to approximations
and assumptions which emphasize the possible similarity of our system to
that of \refs{\bbst}.  In particular, we have treated the string endpoints
on the sevenbranes as if they could be effectively decoupled;
we leave open the problem of treating them more realistically.

$\bullet$ We have not attempted to understand the nonabelian
worldvolume dynamics when $k^\prime > 1$, particularly whether or not
there may be a ``genes' instability'' to fragmentation of the clump.

$\bullet$ A better analysis of the effective dynamics should take into 
account that the lowest modes of the stretched fundamental strings 
are fermionic \bbst.

$\bullet$ It would be interesting to go beyond 
the average-field approximation for the 
charge distribution to see if an electric analog 
of the structure of \lenny\ emerges.

It would be interesting to learn more about this system.

\newsec{What we have to say about biophysics}

Consider the case of two sevenbranes and two D-strings.  
Let the fundamental index of the D-string gauge group 
run over ``purine'' and ``pyrimidine'' $\dots$.

\bigskip
\centerline{\bf{Acknowledgements}}
We thank Savas Dimopoulos, Clifford Johnson, Nemanja Kaloper, Sangmin Lee, 
Lenny Susskind, Scott Thomas and Nick Toumbas for discussions.  
It has been brought to our attention that the title 
of our paper was discovered independently 
\icantbelieveimdoingthis.

\listrefs
\end